\documentclass[aps,prb,twocolumn,groupedaddress,superscriptaddress,showpacs,floatfix,longbibliography]{revtex4-2}
\usepackage{graphicx}
\usepackage{physics}

\usepackage{xcolor}
\usepackage{amsfonts}
\usepackage{amsmath}
\usepackage{amssymb}
\usepackage{rotating}
\usepackage{bm}
\usepackage{bbm}
\usepackage{kantlipsum}
\usepackage{braket}
\usepackage{dsfont}
\usepackage{xtab,afterpage}
\usepackage{lipsum}

\usepackage{makecell}

\def\be{\begin{eqnarray}}
\def\ee{\end{eqnarray}}

\usepackage[unicode=true, bookmarks=false,breaklinks=false,pdfborder={0 0 1},backref=false,colorlinks=false]{hyperref}
\hypersetup{hypertex}   
\usepackage{breakurl}
\hypersetup{backref,colorlinks=true,citecolor=blue,linkcolor=blue,filecolor=blue,urlcolor=blue}

\usepackage[english]{babel}

\begin{document}

\title{
Unified Magnetoelectric Mechanism for Spin Splitting in Magnets}

\author{Carlos Mera Acosta}
\email[]{mera.acosta@ufabc.edu.br}
\affiliation{Center for Natural and Human Sciences, Federal University of ABC, Santo Andre, SP, Brazil}


\begin{abstract}
We identify a magnetoelectric correction that completes the theoretical description of spin splitting (SS) in magnetic systems. Derived from the Dirac equation, this term couples local magnetic moments to the scalar electric potential, providing a third fundamental mechanism—alongside Zeeman and spin–orbit coupling (SOC)—that governs SS in ferromagnets, antiferromagnets, and altermagnets. In compensated magnets, the proposed relativistic correction depends on the difference in electric potential between symmetry-inequivalent motifs, $\mathcal{H}_{\text{ME}} = -\mu_{\text{B}}\eta_0(\mathcal{V}_1 - \mathcal{V}_2)\boldsymbol{\sigma} \cdot \boldsymbol{m}$,
which explains how finite SS emerges in the absence of SOC and enables a complete classification of momentum dependence and motif connectivity across all 32 point groups. Through illustrative examples, we show that distinct SS behaviors—quadratic ($d$-wave altermagnets), linear ($p$-wave altermagnets or spin Zeeman effect), and $k$-independent (SS at $\Gamma$ or fully-compensated ferrimagnets)—are specific manifestations of the proposed magnetoelectric relativistic mechanism, each governed by electric quadrupoles, dipoles, or monopoles, respectively. The formalism naturally extends to higher-order multipoles and more complex symmetries. 
This work establishes a unified framework for SS in magnets and provides a predictive tool for analyzing symmetry-allowed SS in magnetic materials.
\end{abstract}

\maketitle

\section{Introduction}
\begin{figure}[h]
\centering
\includegraphics[width=8.5cm]{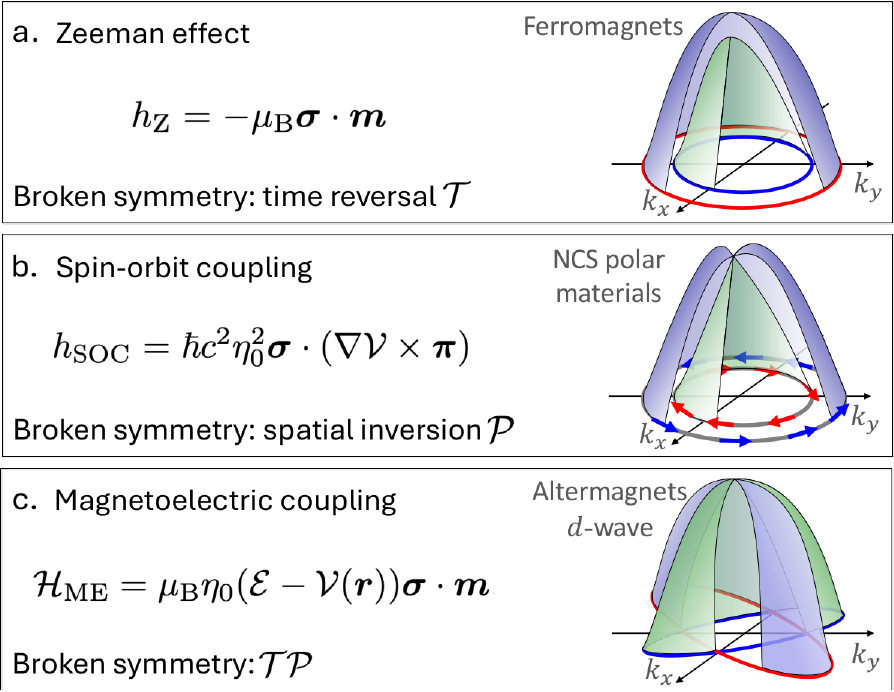}
\label{FigN}
\caption{Comparison of the three fundamental mechanisms of SS in magnetic materials, each linked to a specific relativistic correction and broken symmetry.
(a) Zeeman effect: SS arises from the coupling of spin to the local magnetization $\boldsymbol{m}$, requiring broken time-reversal symmetry $\mathcal{T}$.
(b) SOC: induced by the electric field gradient in noncentrosymmetric (NCS) polar materials, breaking spatial inversion $\mathcal{P}$.
(c) Magnetoelectric coupling: the term $\mathcal{H}_{\text{ME}} = \mu_{\text{B}}\eta_0(\mathcal{E}-\mathcal{V}(\boldsymbol{r}))\boldsymbol{\sigma}\cdot\boldsymbol{m}$ generates SS in systems with broken $\mathcal{T}\mathcal{P}$ symmetry, such as altermagnets. Right panels illustrate representative band structures and spin textures for each case.
}
\end{figure}

Pieter Zeeman’s discovery of the interaction between electron spin and magnetic fields laid the foundation for understanding spin splitting (SS) in magnetic materials~\cite{ZEEMAN1897}.
In the non-relativistic Schrödinger framework, this interaction is introduced \textit{a priori} as $\mu_{\text{B}} g \boldsymbol{\sigma} \cdot \boldsymbol{m}_n$, describing the effect of local magnetic moments $\boldsymbol{m}_n$ that break time-reversal symmetry $\mathcal{T}$ across distinct chemical environments (magnetic motifs $n$).
In ferromagnets, where magnetic motifs are equivalent under translation symmetry $T$, the Zeeman term yields a $k$-independent SS proportional to the Bohr magneton $\mu_{\text{B}}$ and the Landé $g$-factor~\cite{PhysRevLett.62.1560}, as illustrate in Fig.~\ref{FigN}a. Conversely, in compensated magnets, where antiparallel motifs cancel the net magnetization~\cite{PhysRev.87.290}, the same framework predicts the absence of SS unless inversion symmetry $\mathcal{P}$ is also broken.
Specifically, as first envisioned by Emmanuel Rashba, purely electric effects can generate SS (Fig.~\ref{FigN}b): a potential $\mathcal{V}(\boldsymbol{r})$ breaking $\mathcal{P}$ induces a $k$-dependent SS through spin-orbit coupling (SOC)~\cite{bychkov1984properties,arxiv.1812.01721,MERAACOSTA2020145}.

Symmetry analysis and DFT simulations have revealed that $k$-dependent SS can emerge in centrosymmetric compensated magnets when magnetic motifs are not related by either $T$ or $\mathcal{P}$\cite{PhysRevB.102.014422,mejkal2020,Hayami2019,PhysRevX.12.031042,PhysRevB.102.144441,PhysRevB.101.220403,PhysRevB.99.174407,PhysRevMaterials.5.014409,Naka2019,PhysRevB.103.224410,Mazin2021}—that is, in crystals breaking both $\mathcal{S}T$ and $\mathcal{T}\mathcal{P}$ symmetries~\cite{Yuan2023NatComm,Yuan2023,Cheong2025}, where $\mathcal{S}$ denotes spin rotation in $SU(2)$ space. These symmetry-allowed SSs, often termed \textit{non-relativistic} SS, fall into the three categories represented in Figure~\ref{Fig2}: \textit{i}) \textit{Spin Zeeman effect}: a linear magnetoelectric coupling between electric polarization and magnetic moments induces Zeeman-like SS~\cite{PhysRevLett.129.187602}; \textit{ii}) \textit{Altermagnetism}: rotationally connected motifs produce SS only at $k$-points breaking $\mathcal{R}_{n}$ symmetry, with characteristic quadratic forms such as $k_{x}k_{y}$ or $k_x^2 - k_y^2$~\cite{PhysRevX.12.040501}; \textit{iii}) \textit{Non-relativistic SS at $\Gamma$}: when no symmetry relates the motifs, in contrast to altermagnets, a $k$-independent SS arises even at $\boldsymbol{k}=0$~\cite{PhysRevLett.133.216701}. 
A notable subclass of this last category includes fully compensated ferrimagnets, where local magnetic moments—located at chemically distinct atomic environments—cancel exactly to yield zero net magnetization~\cite{PhysRevLett.74.1171}, yet result in finite spin splitting at $\Gamma$~\cite{PhysRevLett.134.116703,Guo2025,arxiv.2412.17232}. 
These effects, though symmetry-permitted, lie beyond conventional Zeeman and Rashba paradigms, raising the fundamental question: what interaction governs SS in compensated magnets in the absence of SOC?

In this work, we identify a relativistic correction that couples the local magnetization $\boldsymbol{m}_n$ to the electric potential $\mathcal{V}n(\boldsymbol{r})$ determined by the local symmetry of magnetic motifs and explains the emergence of SS in compensated magnets without the inclusion of SOC, as illustrate in Fig.~\ref{FigN}.
In compensated magnets, this coupling scales with the difference in electric multipole distributions, $\mu_{\text{B}}(\mathcal{V}_1 - \mathcal{V}_2)\boldsymbol{\sigma} \cdot \boldsymbol{m}$.
The emergence of SS is thus governed by the interplay between magnetic moment orientations, motif symmetries defining the local electric potential, and crystal-imposed motif connectivity. The motif-connectivity as well as the breaking of both $\mathcal{S}T$ and $\mathcal{T}\mathcal{P}$ are naturally obtained from the proposed magnetoelectric correction, which is implemented to explain the SS in compensated magnets with all 32 point group symmetries for magnetic sites and all possible crystal symmetries connecting motifs. 
As illustrative examples, we show that distinct types of SS—altermagnetic, spin Zeeman, and \textit{non-relativistic SS at $\Gamma$}—originate respectively from electric quadrupoles, dipoles, and monopoles in $\mathcal{V}_n$.
The magnetoelectric coupling reveals a $k$-dependence of SS even in ferromagnets, extending beyond the conventional Zeeman effect induced by the exchange field.
Altogether, the magnetoelectric coupling provides a unified framework for understanding SS across compensated magnets, showing that these effects are manifestations of a single physical mechanism.

\section{Derivation of the Magnetoelectric Correction}
The Dirac equation, in contrast to the non-relativistic Schrödinger framework, encodes spin from first principles, unveiling Zeeman- and SOC-induced SS as intrinsic relativistic phenomena~\cite{1928}.
Decomposing the four-component spinor into large ($\chi_{+}$) and small ($\chi_{-}$) components, the Dirac formalism yields the coupled equations $c\boldsymbol{\sigma}\cdot\boldsymbol{\pi}\chi_{\mp} = (E-\mathcal{V}(\boldsymbol{r})\mp mc^{2})\chi_{\pm}$, equivalently, for the large component (with $\mathcal{E} = E - mc^{2}$), \begin{equation} c^{2}\boldsymbol{\sigma}\cdot\boldsymbol{\pi} (2mc^{2} + \mathcal{E} - \mathcal{V}(\boldsymbol{r}))^{-1} \boldsymbol{\sigma}\cdot\boldsymbol{\pi}\chi_{+} = (\mathcal{E} - \mathcal{V}(\boldsymbol{r}))\chi_{+}. \end{equation} The covariant momentum operator $\boldsymbol{\pi} = \boldsymbol{p} - e\boldsymbol{A}$ accounts for minimal coupling to the magnetic vector potential $\boldsymbol{A}$ associated with local magnetization, while $\mathcal{P}$-breaking is encoded in the electric potential $\mathcal{V}(\boldsymbol{r}) = e\phi$.
the Dirac formalism can be simplified by expanding 
\begin{equation}
(2mc^{2} + \mathcal{E} - \mathcal{V}(\boldsymbol{r}))^{-1} = \eta_{0} + \eta_{1} + \dots,\end{equation}
with $\eta_{0} = 1/2mc^{2}$ and $\eta_{1} = -\eta_{0}^{2}(\mathcal{E} - \mathcal{V}(\boldsymbol{r}))$. 
The zero-order relativistic correction, $c^{2}(\boldsymbol{\sigma}\cdot\boldsymbol{\pi})\eta_{0}(\boldsymbol{\sigma}\cdot\boldsymbol{\pi})$, yields the Pauli equation \begin{equation}
(\boldsymbol{\pi}^{2}/2m + h_{\text{Z}})\chi_{+} = (\mathcal{E} - \mathcal{V}(\boldsymbol{r}))\chi_{+},\end{equation} where the Zeeman term $h_{\text{Z}}=-\mu_{\text{B}}\boldsymbol{\sigma}\cdot\boldsymbol{m}$ arises from the non-commutativity of the covariant momenta, $[\pi_{i}, \pi_{j}] = -ie\hbar \epsilon_{ijk} m_{k}$. 
In contrast, the first-order correction 
\begin{equation}
c^{2}(\boldsymbol{\sigma}\cdot\boldsymbol{\pi})\eta_{1}(\boldsymbol{\sigma}\cdot\boldsymbol{\pi})=-c^{2}(\boldsymbol{\sigma}\cdot\boldsymbol{\pi})\eta^{2}_{0}(\mathcal{E} - \mathcal{V}(\boldsymbol{r}))(\boldsymbol{\sigma}\cdot\boldsymbol{\pi}),\end{equation} 
includes the well established conventional SOC, $c^{2}\eta_{0}^{2}\hbar\boldsymbol{\sigma}\cdot(\nabla\mathcal{V}(\boldsymbol{r})\times\boldsymbol{\pi})$,
the kinetic energy relativistic correction, $c^{2}\eta_{0}^{2}(\mathcal{E} - \mathcal{V}(\boldsymbol{r}))\boldsymbol{\pi}^{2}$, the Darwin term, $ic^{2}\eta_{0}^{2}\hbar\nabla\mathcal{V}(\boldsymbol{r})\cdot\boldsymbol{\pi}$, 
and a previously unnoticed term we identify as a magnetoelectrical relativistic correction (Fig. \ref{FigN}c), 
\begin{equation}\mathcal{H}_{\text{ME}} = \mu_{\text{B}}\eta_{0}(\mathcal{E} - \mathcal{V}(\boldsymbol{r}))\boldsymbol{\sigma}\cdot\boldsymbol{m}.\end{equation}
Magnetic and electric relativistic corrections thus emerge at different orders (Zeeman at zeroth, SOC at first) often justifying independent treatments, yet overlooking the intrinsic magnetoelectric coupling unveiled here.

\begin{figure}[h]
\centering
\includegraphics[width=8.5cm]{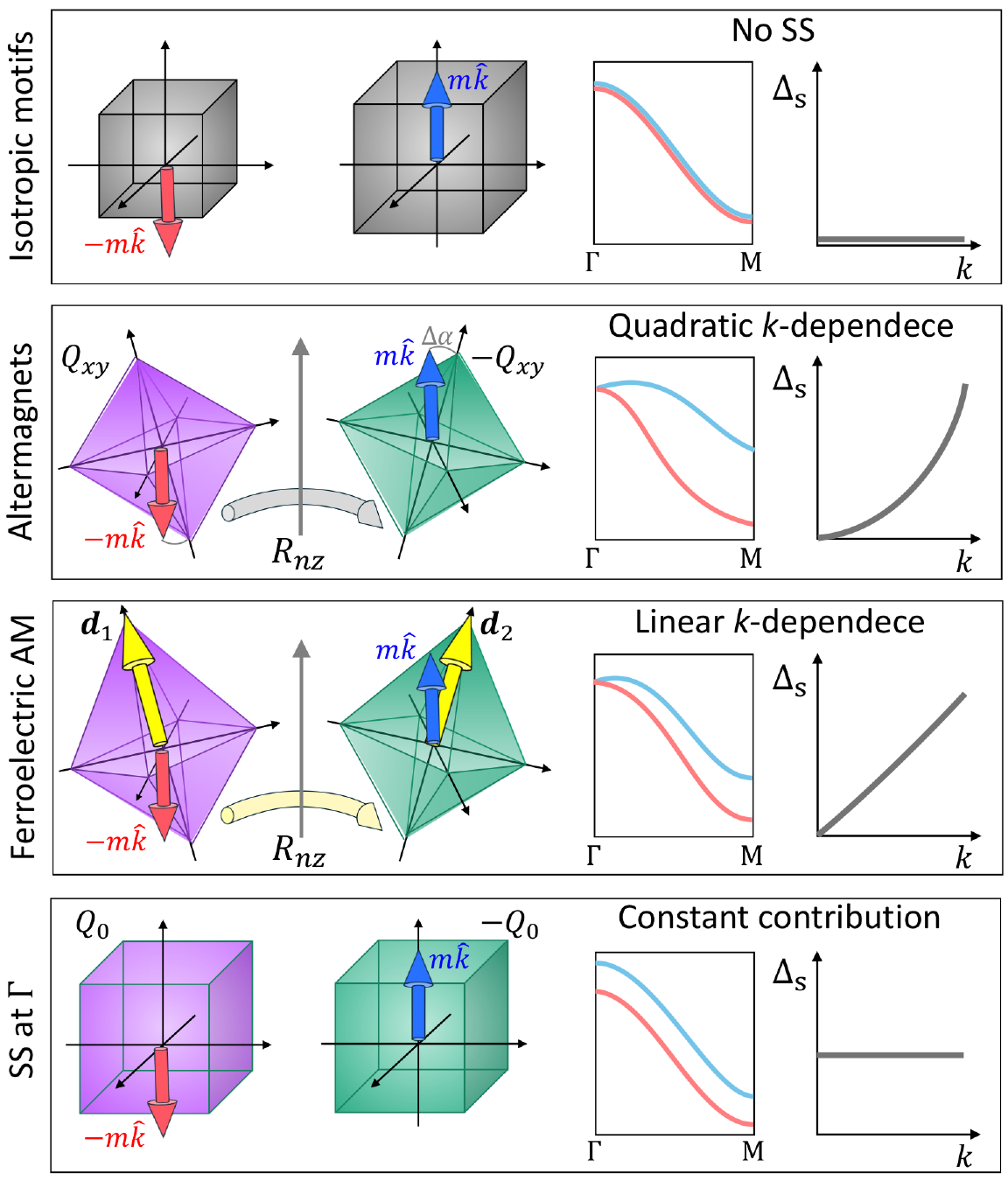}
\label{Fig2}
\caption{Illustration of magnetic motifs with antiparallel magnetic moments, schematic representative SS behavior from the $\Gamma$ point to an arbitrary symmetry point M, and SS ($\Delta_{\text{s}}$) $k$-dependence for isotropic motifs, $d$-wave altermagnets (AM), ferroelectric altermagnets (dipoles are shown in yellow), and AFM SS at the $\Gamma$ points. Difference on the local electric potential in motifs are represented by purple and green motifs for quadrupole $\mathcal{Q}_{xy}$, dipole $\boldsymbol{d}$, and charge $Q_{0}$. 
}
\end{figure}


To understand the origin of the magnetoelectric term explaning the emergence of SS in compensated magnets, 
we note that the electric potential does not commute with the covariant momentum operator, 
in contrast to the textbook approach in which $\eta_{1}$ is consider as a constant. We thus examine the first-order correction, treating $\eta^{2}_{0}(\mathcal{E} - \mathcal{V}(\boldsymbol{r}))$ as a single operator entity, and hence preserving all nontrivial contributions from electric gradients and potential-dependent commutators. 
Applying the identity
$
\left[ \pi_i, \mathcal{V}(\boldsymbol{r}) \right] = -i\hbar \partial_i \mathcal{V}$,
and using the linearity of the Pauli matrix contraction, we obtain
\begin{equation}
    \left( \boldsymbol{\sigma} \cdot \boldsymbol{\pi} \right)
(\mathcal{E} - \mathcal{V})
= (\mathcal{E} - \mathcal{V})
\left( \boldsymbol{\sigma} \cdot \boldsymbol{\pi} \right)
- i\hbar \boldsymbol{\sigma} \cdot \nabla \mathcal{V}, 
\end{equation}
which allow us to interpret the first-order relativistic correction as two distinct terms: one in which the gradient of the electric potential appears explicitly and another in which the potential factor multiplies the square of the Pauli-covariant product--- analogous to the zero-order correction leading to the Zeeman effect in the Pauli equation, i.e., 
\begin{equation}
-c^{2}\eta^{2}_{0}[ -i\hbar (\boldsymbol{\sigma} \cdot \boldsymbol{\nabla} \mathcal{V})
\left( \boldsymbol{\sigma} \cdot \boldsymbol{\pi} \right) + (\mathcal{E} - \mathcal{V})
\left( \boldsymbol{\sigma} \cdot \boldsymbol{\pi} \right)^2 ].
\end{equation}

To evaluate the first term, the operator product
$i\hbar\boldsymbol{\sigma} \cdot (\nabla \mathcal{V}) \;
\boldsymbol{\sigma} \cdot \boldsymbol{\pi}$
is rewritten using the Pauli matrix identity as $i\hbar\boldsymbol{\nabla} \mathcal{V} \cdot \boldsymbol{\pi}
-\hbar\boldsymbol{\sigma} \cdot (\boldsymbol{\nabla} \mathcal{V} \times \boldsymbol{\pi})$. Here, $i\hbar\boldsymbol{\nabla} \mathcal{V} \cdot \boldsymbol{\pi}$ is not manifestly Hermitian. Upon symmetrization—adding its Hermitian conjugate—we obtain the so-called Darwin term, $h_{\mathrm{D}} = -\frac{\hbar^2}{8m^2 c^2} \nabla^2 \mathcal{V}$,
which represents the zitterbewegung smearing of the electron position due to rapid spinor component oscillations, and appears as a scalar correction to the effective potential~\cite{PhysRevB.72.085217}. The term $-\hbar\boldsymbol{\sigma} \cdot (\boldsymbol{\nabla} \mathcal{V} \times \boldsymbol{\pi})$, being already Hermitian, yields the conventional SOC term,
$h_{\mathrm{SOC}} = \hbar c^{2}\eta_{0}^{2}\boldsymbol{\sigma} \cdot (\nabla \mathcal{V} \times \boldsymbol{\pi})$. 
The SOC-induced SS vanishes in inversion-symmetric environments where the electric potential has no gradient, but becomes important in materials with broken inversion symmetry, such as polar semiconductors or surfaces~\cite{PhysRevB.102.144106,PhysRevB.94.041302,PhysRevLett.122.036401}.

On the other hand, in the second term in the first-order relativistic correction, i.e., $(\mathcal{E} - \mathcal{V})
\left( \boldsymbol{\sigma} \cdot \boldsymbol{\pi} \right)^2$, we expand the square of the Pauli product using the standard identity and the non-commutativity of the covariant momenta, resulting in $(\mathcal{E} - \mathcal{V})[\boldsymbol{\pi}^2 - e\hbar \boldsymbol{\sigma} \cdot \boldsymbol{m}]$. 
Consequently, we obtain a relativistic correction to the kinetic energy
$h_{\mathrm{k}} = -c^{2}\eta_{0}^{2}(\mathcal{E} - \mathcal{V}) \boldsymbol{\pi}^2$, and one remaining term proportional to $(\mathcal{E} - \mathcal{V})( -e\hbar \boldsymbol{\sigma} \cdot \boldsymbol{m})$, which corresponds to the magnetoelectric correction $\mathcal{H}_{\text{ME}}=\mu_{\text{B}}\eta_{0}(\mathcal{E} - \mathcal{V}(\boldsymbol{r}))\boldsymbol{\sigma}\cdot\boldsymbol{m}$. 
This term modifies the usual spin-magnetic coupling by weighting it with the electric potential, thereby introducing a spin-dependent energy shift that depends on both the magnetic texture and the electrostatic environment. 

Each of relativistic terms stems from a different relativistic origin: the Zeeman and magnetoelectric couplings result from the magnetic field and its interaction with the spinor structure, while SOC and Darwin terms arise from electric gradients, and the kinetic term renormalizes the nonrelativistic dispersion. 
This hierarchy of relativistic corrections explains why SS persists even in systems with negligible SOC, as long as local electric fields differ across magnetic motifs, as we discuss below. Crucially, the magnetoelectric correction remains finite even when SOC-induced splitting vanishes by symmetry or is explicitly excluded. 
Note that for simplicity, and without loss of generality, we have assumed throughout the deduction of the Zeeman and magnetoelectric terms that the magnetic field enters as linearly proportional to the local magnetic moment. This approximation captures the leading-order effects responsible for SS and allows for a transparent analytical treatment of the magnetoelectric coupling mechanism. It should be noted, however, that the full magnetic field associated to the magnetic vector potential $\boldsymbol{A}(\boldsymbol{r})$
arises from all magnetic sources in the crystal, including higher-order magnetic multipoles such as magnetic quadrupoles or toroidal moments. These additional contributions may introduce spin-dependent corrections with higher-order $\boldsymbol{k}$-dependence, particularly in systems with noncollinear order or spatially inhomogeneous magnetic textures. While such terms are formally present in the complete relativistic expansion, their amplitudes are typically small compared to the dipolar component. 
Consequently, for the magnetic dipole contributions, the full effective Hamiltonian to order $c^{-2}$ reads
\begin{equation}
    \mathcal{H}_{\mathrm{eff}} =
\frac{\boldsymbol{\pi}^2}{2m}
+ h_{\text{Z}}
+ \mathcal{H}_{\mathrm{ME}} + h_{\mathrm{SOC}} + h_{\mathrm{D}} + h_{\mathrm{k}} + \mathcal{O}(c^{-4}),
\end{equation}
where the contributions inducing SS can be interpreted as on-site terms. For instance, the magnetoelectric correction at a given magnetic motif $n$ characterized by an electric potential $\mathcal{V}_n(\boldsymbol{r})$ and magnetic moment $\boldsymbol{m}_n$, is proportional to the product $\mathcal{V}_n(\boldsymbol{r})\boldsymbol{\sigma}\cdot\boldsymbol{m}_n$.

In general, in an arbitrary magnet, the resulting SS ($\Delta_{\text{ss}}$) stems from the combined effects of the Zeeman term, SOC, and magnetoelectric term, captured by
\begin{equation}
\Delta_{\text{ss}}=-\mu_{\text{B}}\boldsymbol{\sigma}\cdot[\boldsymbol{m}_n + (\eta_0/e)(\nabla \mathcal{V}_n \times \boldsymbol{\pi}_n) - \eta_0(\mathcal{E}-\mathcal{V}_n)\boldsymbol{m}_n],
\end{equation}
which is valid for both collinear and non-collinear magnetic configurations. 
This insight justifies that the magnetoelectric mechanism is a third mechanism for SS across magnetic systems, consistent with relativistic corrections and crystal symmetry. 
The proposed magnetoelectric mechanism confirms that a crucial ingredient controlling SS is the motif symmetry, which dictates the presence or absence of specific multipolar tensors in the expansion of the electric potential~\cite{PhysRevB.98.245129,Hayami2018,PhysRevB.98.165110},  
\begin{equation}
\mathcal{V}_n(\boldsymbol{r}) = \lambda_0 Q_n + \lambda_1 \sum_i d_{ni} \hat{r}_i + \lambda_2 \sum_{ij} \mathcal{Q}_{nij} \hat{r}_i \hat{r}_j + \dots,
\end{equation}
where $\lambda_l = (1/4\pi\varepsilon_0)\langle r^{-(l+1)} \rangle$ defines the amplitude of each $l$-rank multipolar tensor: electric charge $Q_n$ ($l=0$), dipoles $\boldsymbol{d}_n$ ($l=1$), and quadrupoles $\mathcal{Q}_n$ ($l=2$). 
Once the local motif symmetry is fixed, a crystal symmetry operation $\mathcal{U}$ determines the magnetic motifs connectivity, i.e., in a unit cell with two motifs, their electric dipoles and quadrupoles relate via $\boldsymbol{d}_{2} = \mathcal{U} \boldsymbol{d}_{1}$ and $\mathcal{Q}_{2} = \mathcal{U} \mathcal{Q}_{1} \mathcal{U}^{\dagger}$. 

In magnets, the resulting SS arises from the interplay between magnetic ordering, site symmetry, and motif connectivity. 
For example, for antiparallel magnetic moments ($\boldsymbol{m}_{n} = (-1)^{n-1} \boldsymbol{m}$ with $n=1,2$), the Zeeman terms cancel and the magnetoelectric contribution is
\begin{equation}
\mathcal{H}_{\text{ME}} = -\mu_{\text{B}}\eta_{0}(\mathcal{V}_{1} - \mathcal{V}_{2})\boldsymbol{\sigma}\cdot\boldsymbol{m}.
\end{equation}
This spatial modulation of the electric potential provides a microscopic mechanism that substantiates the phenomenological models proposed by Rashba and Pekar~\cite{Pekar1965Combined} and Gomonay \textit{et al.}~\cite{Gomonay2024} based on spatial inhomogeneities in magnetic moments. 
Electric potential differences ($\mathcal{V}_{1}-\mathcal{V}_{2}$) can arise from external electric fields, as confirmed by DFT calculations in 2D monolayers~\cite{PhysRevB.108.L180403}, or intrinsically, from anisotropies in the local crystallographic environment and orbital ordering~\cite{PhysRevLett.132.236701}.

Importantly, this magnetoelectric term $\mathcal{H}_{\text{ME}}$ is symmetry-allowed precisely under the same conditions previously identified as necessary for SS in compensated magnets, namely, when both $\mathcal{S}T$ and $\mathcal{T}\mathcal{P}$ symmetries are broken. 
This reveals that $\mathcal{H}_{\text{ME}}$ provides the microscopic mechanism responsible for the SS emergence under the symmetry constraints established in prior group-theoretical classifications.
As discussed in the following, the nature of the electric multipole that governs $\mathcal{V}_n$ (quadrupole, dipole, or monopole) dictates the emergence of altermagnetic SS, spin Zeeman SS or SS at $\Gamma$, respectively, all encoded by motif symmetry and connectivity, as represented in Figure~\ref{Fig2}. 

It is important to emphasize, however, that the identification of a dominant multipole—monopole, dipole, or quadrupole—does not imply the absence of others. By symmetry, the presence of a given electric multipole generally permits the existence of higher-order terms. For instance, a site supporting a monopolar contribution ($Q_n \neq 0$) also admits dipolar, quadrupolar, and higher-rank moments in its potential expansion. Likewise, dipolar motifs typically host quadrupolar and octupolar components, albeit with diminishing amplitude and influence. While our illustrative examples focus on dominant contributions for clarity, all symmetry-allowed multipoles are, in principle, present and contribute to the total SS. 
Nevertheless, the framework we present is general and can be straightforwardly extended to include higher-rank contributions and more complex motif symmetries. 

\section{Quadrupole-Induced Spin Splitting in Altermagnets}
Table~I summarizes the point group conditions and canonical forms of the allowed quadrupole tensor components $\mathcal{Q}_{ij}$: 
\textit{(i)} In isotropic environments, all tensor elements vanish, $\mathcal{Q}_{ij} = 0$; 
\textit{(ii)} In axial environments, the tensor adopts a uniaxial diagonal form with $\mathcal{Q}_{xx} = \mathcal{Q}_{yy}$ and $\mathcal{Q}_{zz} = -2\mathcal{Q}_{xx}$, forbidding all off-diagonal terms;
\textit{(iii)} In lower-symmetry motifs with a single mirror or improper rotation, at least one off-diagonal component survives depending on the mirror orientation, e.g., $\mathcal{Q}_{xy} \neq 0$;
\textit{(iv)} In fully unconstrained sites, all six independent components of $\mathcal{Q}_{ij}$ may be nonzero. 
The impact of motif connectivity is captured by a crystal symmetry operation $\mathcal{U}$ transforming the quadrupole tensor as $\mathcal{Q}_2 = \mathcal{U} \mathcal{Q}_1 \mathcal{U}^\dagger$, resulting in the magnetoelectric correction
\begin{equation}
    \mathcal{H}_{\text{ME}} = -\mu_{\text{B}} \eta_{0}\lambda_{2} \boldsymbol{\sigma}\cdot\boldsymbol{m} \sum_{ij} \left( \mathcal{Q}_{1ij} - [\mathcal{U} \mathcal{Q}_1 \mathcal{U}^\dagger]_{ij} \right) \hat{r}_i \hat{r}_j.
\end{equation}
If $\mathcal{U}$ \emph{commutes} with the canonical form of $\mathcal{Q}_1$, the magnetoelectric correction vanishes and spin degeneracy is preserved, i.e., $[\mathcal{U}, \mathcal{Q}_1]=0$ (see Table~I). 
Breaking such symmetries through motif connectivity yields finite SS.
For instance, since $\mathcal{P}$ always commutes with $\mathcal{Q}_1$, breaking inversion-related motif connectivity suffices to induce SS, even between polar NCS motifs.
Moreover, if $\mathcal{U}$ belongs to the motif symmetry itself, its breaking through connectivity also modifies the motif, leading to $\mathcal{Q}_2 \neq \mathcal{U} \mathcal{Q}_1 \mathcal{U}^\dagger$.
These results align with Yuan \textit{et al.}~\cite{PhysRevB.102.014422}, who showed that SS is forbidden when motifs are connected by $\mathcal{P}$ and $T$ symmetries.
Notably, in isotropic environments ($\mathcal{Q}_1=0$), spin degeneracy persists even without inversion-related motif connectivity, implying that SS additionally demands site symmetries permitting nonzero quadrupoles (first line in Figure~\ref{Fig2}). 
This implies that in the quadrupole regime, even if both $\mathcal{T}\mathcal{P}$ and $\mathcal{S}T$ are broken, there is no SS. However, the local motif symmetry can allow higher multipoles.


When crystal operations $\mathcal{U}$ \emph{fails to commute} with the quadrupole tensor, i.e., $[\mathcal{U}, \mathcal{Q}_1] \neq 0$, the resulting difference $\Delta \mathcal{Q} = \mathcal{Q}_1 - \mathcal{U} \mathcal{Q}_1 \mathcal{U}^\dagger$ becomes finite, leading to a non-zero magnetoelectric term and thus to a finite SS.
For nonisotropic motifs, such noncommuting operations always exist; in particular, rotation $R_n$ and improper rotation $S_n$ symmetries are sufficient to guarantee SS (second line in Figure~\ref{Fig2}). 
These considerations align with the design principles proposed for altermagnetism~\cite{PhysRevX.12.031042}. 
Note that while Yuan \textit{et al.} analyze the symmetry constraints preserving spin degeneracy~\cite{PhysRevB.102.014422}, Smejkal \textit{et al.} emphasize the symmetry-breaking mechanisms that enable SS~\cite{PhysRevX.12.031042}. We reconcile both perspectives, showing that they present complementary views of the same magnetoelectric mechanism identified here. 

In the magnetoelectric SS (equation 5), crystal operation $\mathcal{U}$ dictates the $k$-dependence.  
This is exemplified by MnF$_2$ (P$4_2$/mnm)~\cite{PhysRevB.102.014422}, a prototypical altermagnet featuring two chemically distinct Mn sites, Mn$_n$ ($n=1,2$), each with local $C_{2h}$ symmetry and quadrupole
\begin{equation}
\mathcal{Q}_n =
\begin{pmatrix}
\mathcal{Q}_{x} & (-1)^{n-1}\mathcal{Q}_{xy} & 0 \\
(-1)^{n-1}\mathcal{Q}_{xy} & \mathcal{Q}_{x} & 0 \\
0 & 0 & -2\mathcal{Q}_{x}
\end{pmatrix},
\end{equation}
where $\mathcal{Q}_x$ and $\mathcal{Q}_{xy}$ reflect local electrostatics and atomic coordination. 
Bader analysis at room temperature yields $\mathcal{Q}_{x} \approx 1.31 \times 10^{-2}$ and $\mathcal{Q}_{xy} \approx 1.39 \times 10^{-2}$ a.u~\cite{PhysRevX.14.011019}.
The motifs are connected by a rotation $R_{4c'}(\phi)$ ($\phi = 2\pi/4$) around the principal axis $\boldsymbol{c}'$, perpendicular to the mirror plane $\sigma_h$.
This operation maps $\mathcal{Q}_1$ onto $\mathcal{Q}_2 = R_{4c'}\, \mathcal{Q}_1\, R_{4c'}^\dagger$, flipping the sign of $Q_{xy}$ while preserving $Q_x$.
Thus, in the magnetoelectric difference $\mathcal{Q}_1 - \mathcal{Q}_2$, diagonal components cancel while off-diagonal ones add constructively.
The resulting SS is governed by
\begin{equation}
\mathcal{H}_{\text{ME}}(\boldsymbol{k}) = -2\mu_{\text{B}} \eta_0 \lambda_2 \boldsymbol{\sigma}\cdot\boldsymbol{m}\, Q_{xy} k_x k_y,
\end{equation}
capturing the characteristic anisotropic, quadratic $k$-dependence expected in the altermagnetic SS~\cite{PhysRevB.110.144412}.

\onecolumngrid
\begin{table*}[htb]
\centering
\caption{Electric quadrupole tensor constraints across the crystallographic point groups classified according to symmetry classes. 
An arbitrary axis defined by the orthogonal vectors $\boldsymbol{a}$, $\boldsymbol{b}$, and $\boldsymbol{c}$ is considered, where the axial symmetry is along $\boldsymbol{c}$. 
Motif symmetries are  divided into centrosymmetric (CS), non-centrosymmetric (NCS) nonpolar, and polar point groups according to the quadrupole tensor canonical form. 
Symmetry operations that \emph{commute} with the canonical form of the quadrupole tensor $\mathcal{Q}$ and those that \emph{do not} (generating a finite difference $\Delta\mathcal{Q} = \mathcal{Q}_{1} - \mathcal{U} \mathcal{Q}_{1}\mathcal{U}^\dagger$) are identified.
The notation $\mathcal{Q}_{c}=\mathcal{Q}_{x} + \mathcal{Q}_{b}$ and $\mathcal{Q}_{i}=\mathcal{Q}_{ii}$ is adopted.
}
\renewcommand{\arraystretch}{1.15}
\begin{tabular}{c c c c c c c c}

\hline
\makecell{Symmetry\\classes} & Canonical $Q$ & \makecell{CS} & \makecell{NCS\\Nonpolar} & \makecell{NCS\\ Polar } & \makecell{No SS: $\mathcal{U}$ for\\$[\mathcal{U},\mathcal{Q}]=0$} & \makecell{Magnetoelectric SS:\\$\mathcal{U}$ for $[\mathcal{U},\mathcal{Q}]\neq0$} & \makecell{Example of\\$k$-dependence} \\

\hline
Isotropic & $\mathcal{Q}_{ij}=0$ & $T_h$, $O_h$ & $T$, $O$, $T_d$ & --- & \makecell{All $\mathcal{U}$} & \makecell[c]{None} & --- \\
\\

\makecell{Uniaxial \\ $n\geq 3$}& \makecell{ 
$\begin{pmatrix}
\mathcal{Q}_{a} & 0 & 0\\
0 & \mathcal{Q}_{a} & 0\\
0 & 0 & -2\mathcal{Q}_{a}
\end{pmatrix}$} & \makecell{$C_{4h}$,$D_{4h}$\\$C_{6h}$,$D_{6h}$\\$C_{3i}$,$D_{3d}$} & \makecell{S$_4$,$D_{2d}$\\$C_{3h}$,$D_{3h}$\\$D_{n}$} &
\makecell{$C_n$\\$C_{nv}$} & \makecell{$E$, $\mathcal{P}$\\$R_{n}\parallel\boldsymbol{c}$\\ $S_{n}\parallel\boldsymbol{c}$\\ $\sigma_{v}\parallel\boldsymbol{a}$ or $\boldsymbol{b}$} & \makecell[c]{$R_{n}\perp\boldsymbol{c}$, $n\neq2$\\
$S_{n}\perp\boldsymbol{c}$, $n\neq2$\\
Any axis‑tilting $\mathcal{U}$} & \makecell{For $R_{n}\perp\boldsymbol{c}$,\\ with $R_{n}\parallel\boldsymbol{a}$ and $n=4$:\\$3\mathcal{Q}_{a}(k^{2}_{b}-k^{2}_{c})$}\\
\\

\makecell{Uniaxial \\ $n=2$} &
\makecell{$\begin{pmatrix}
\mathcal{Q}_{a} & 0 & 0\\
0 & \mathcal{Q}_{b} & 0\\
0 & 0 & -\mathcal{Q}_{c}
\end{pmatrix}$} &$D_{2h}$ &$D_2$ & \makecell{$C_n$\\$C_{nv}$} &
\makecell{$E$, $\mathcal{P}$, $R_{2c}$\\$R_{2a}$, $R_{2b}$\\$\sigma_a$, $\sigma_b$, $\sigma_c$} & \makecell[c]{$R_{m}\parallel\boldsymbol{c}$, $m\neq2$\\
$R_{n}$ in diagonal axes\\
$S_{n}\parallel$ or $S_{n}\perp\boldsymbol{c}$, $n\neq2$\\
$\sigma_{d}$ bisecting $\boldsymbol{a}$ and $\boldsymbol{b}$\\
Any axis‑tilting $\mathcal{U}$} & \makecell{For $R_{m}\parallel\boldsymbol{c}$,\\ with $m=4$:\\$(\mathcal{Q}_{a}-\mathcal{Q}_{b})(k^{2}_{a}-k^{2}_{b})$}\\
\\

\makecell{Single mirror \\ $\sigma$} & \makecell{At least one off-diag.:\\$\begin{pmatrix}
\mathcal{Q}_{a} & \mathcal{Q}_{ab} & 0\\
\mathcal{Q}_{ab} & \mathcal{Q}_{b} & 0\\
0 & 0 & -\mathcal{Q}_{c}
\end{pmatrix}$} & $C_{2h}$ & --- & $C_s$ & \makecell{$E$, $\mathcal{P}$, $\sigma$ \\ ($R_{2c}$ in $C_{2h}$)} & \makecell[c]{$R_{n}\parallel$ or $R_{n}\perp\sigma$, $n\neq2$\\
$S_{n}\parallel$ or $S_{n}\perp\sigma$, $n\neq2$\\
Orthogonal mirrors $\sigma_{\perp}$\\
Any $\mathcal{U}$ that tilts $\sigma$} & \makecell{For $R_{n}\perp\sigma$,\\ with $n=4$:\\$2\mathcal{Q}_{ab}k_{a}k_{b}$}\\
\\

\makecell{Unrestricted} & \makecell{$\mathcal{Q}_{ij}\neq0$} & $C_i$ & --- & $C_1$ & \makecell{$E$, $\mathcal{P}$} & \makecell[c]{$R_{n}$, $\forall n\ge2$, any axis\\
$S_{n}$, $\forall n\ge2$, any axis\\
All mirrors $\sigma$} & \makecell{For $R_{n}\parallel\boldsymbol{c}$, \\ with $n=2$:\\$2\mathcal{Q}_{ac}k_{a}k_{c}+2\mathcal{Q}_{bc}k_{b}k_{c}$}\\

\hline
\end{tabular}
\label{tab:GenericRotationQuadrupole}
\end{table*}
\twocolumngrid

As previously noted, in centrosymmetric neutral motifs, charge and dipole moments are symmetry-forbidden, making the quadrupole the leading allowed multipole. 
However, the dominance of the quadrupolar contribution to SS does not preclude the presence of higher-order multipoles. Such contributions can yield quartic or sextic $k$-dependencies, characteristic of $g$-wave and $i$-wave altermagnets~\cite{PhysRevX.12.031042}. This is also valid for magnetic multipoles.  
Magnetic octupoles also trace the symmetry of altermagnetic SS~\cite{PhysRevX.14.011019}, but their identified SOC dependence contrasts with the SOC-independence of the SS itself, a distinction we will address elsewhere. 
The $k$-dependent nature of magnetic octupoles also precludes them from accounting for SS at the $\Gamma$ point. 
Interestingly, the quadrupole components in the altermagnetic MnF$_2$ have been shown to be independent of SOC~\cite{PhysRevX.14.011019}. Moreover, the intrinsic quadrupole symmetry matches that of the symmetric second-rank tensors introduced phenomenologically to describe altermagnets~\cite{Gomonay2024}. 

The relevance of quadrupolar interactions is further supported by experimental evidence in UNi$_4$B, where coupling between quadrupoles and lattice strain manifests as elastic softening in both antiferromagnetic and paramagnetic phases~\cite{PhysRevLett.126.157201}. 
From this perspective, altermagnets emerge as a distinct class of multiferroic systems, where electric and magnetic orders coexist. However, unlike conventional multiferroics, where ferroelectricity and magnetism are intrinsically coupled~\cite{PhysRevLett.96.067601,PhysRevLett.96.097202,PhysRevB.76.144424,Mostovoy2024},  
the electric and magnetic multipoles in altermagnets may originate independently, yet still combine to produce SS through the local relativistic coupling $\mathcal{H}_{\text{ME}}$.

\section{Dipole-Driven Spin Zeeman Effect in Compensated Magnets}  
Having established the quadrupole-mediated mechanism underlying altermagnetic SS, we now address the dipolar regime.  
Specifically, we examine how electric dipoles combine with local magnetic moments to generate a linear-in-$k$ SS, classified as the \textit{spin Zeeman effect}~\cite{PhysRevLett.129.187602} or electric potential difference antiferromagnetism~\cite{PhysRevB.108.L180403} when driven by external electric fields.  
Electric dipoles $d_i$ are symmetry-allowed in polar environments ($C_n$ and $C_{nv}$), but forbidden at CS and NCS nonpolar sites, as summarized in Table~I.  
Crystal symmetry further constrains motif connectivity: in NCS crystals, the absence of inversion symmetry allows only polar or nonpolar sites; in CS crystals, polar and nonpolar motifs must occur in inversion-related pairs, canceling their dipole contributions, particularly when local symmetries match.  
Conversely, CS sites may combine freely under inversion.  

\begin{table}[htb]
\centering
\caption{Classification of symmetry operations at polar sites based on their action on the local electric dipole $\boldsymbol{d}\parallel \boldsymbol{c}$. 
Operations are grouped into three classes: those preserving $\boldsymbol{d}$, those inverting it, and those reorienting it into $\boldsymbol{d}'\neq\pm\boldsymbol{d}$.
}

\begin{tabular}{c c c}
\hline
\makecell{Symmetry classes} &
\makecell{Symmetries\\ connecting motifs }&
\makecell{Example of\\ $k$-dependence} \\ \hline

\makecell{Preserving\\$\boldsymbol{d}\rightarrow \boldsymbol{d}$} & \makecell[c]{$E$, $\sigma_{v}\parallel\boldsymbol{c}$ \\ $R_{n}\parallel \boldsymbol{c}$ $\forall n$} & \makecell{None} \\
\\
\makecell{Inverting\\$\boldsymbol{d}\rightarrow -\boldsymbol{d}$} & \makecell[c]{$\mathcal{P}$, $\sigma_{h}$($xy$)\\$R_{2}\perp \boldsymbol{c}$, $S_{n}\parallel \boldsymbol{c}$}& \makecell{For $\mathcal{P}$-related motifs:\\ $\boldsymbol{d}\cdot\boldsymbol{k}$}\\
\\
\makecell{Reorienting\\$\boldsymbol{d}\rightarrow \boldsymbol{d}^{\prime}$ } &
\makecell[c]{Obliques $R_{n}$, $S_{n}$\\ and diagonal $\sigma_{d}$} & \makecell{For $R_{n}$ with $n=2$:\\ $d_{x}k_{x}+d_{y}k_{y}$}\\ 
\hline
\end{tabular}
\label{tab:AxisRelationPolar}
\end{table}

In the trivial case of a compensated CS magnet composed of two isotropic CS motifs, both dipole and quadrupole contributions vanish, resulting in no SS. For non-isotropic motifs in CS or NCS nonpolar environments, where electric dipoles are symmetry-forbidden, SS arises solely from differences in the quadrupole tensor. In contrast, when polar magnetic motifs are present in NCS or CS crystals and are connected by a crystal symmetry operation $\mathcal{U}$, the magnetoelectric coupling becomes
\begin{equation}
\mathcal{H}_{\text{ME}} = -\mu_{\text{B}} \eta_{0}\lambda_{1} \boldsymbol{\sigma}\cdot\boldsymbol{m} \sum_{i} \left( d_{1,i} - [\mathcal{U} d_1]_{i} \right) \hat{r}_i.
\end{equation}
Here, the operation $\mathcal{U}$ may preserve, invert, or reorient the dipole direction, as classified in Table~II.

In ferroelectric altermagnets, where local motifs are connected by crystal rotational symmetries aligned with their local polar axis (see Table~II), the electric dipole is preserved across sites, causing the magnetoelectric contribution from dipoles to vanish. 
Consequently, SS in these systems arises exclusively from electric quadrupoles. 
This behavior contrasts sharply with ferroelectric Rashba systems~\cite{PhysRevB.102.144106,fphy.2014.00010,Rinaldi2018}, where spin polarization (SP) couples directly to electric dipoles and reverses upon polarization switching. 
In ferroelectric altermagnets, by contrast, reversing the electric polarization does not necessarily invert the SP. 
Indeed, as reported by Gu et. al.~\cite{PhysRevLett.134.106802}, most such systems exhibit robust SP direction even under dipole reversal, highlighting the quadrupolar (not dipolar) origin of the SS.

In contrast, when motif connectivity enforces antiparallel alignment of local dipoles, as in antiferroelectric altermagnets~\cite{PhysRevLett.134.106801}, the dipolar contributions add constructively in the magnetoelectric term, leading to a linear-in-$\boldsymbol{k}$ SS consistent with the \textit{spin Zeeman effect} proposed by Zhao \textit{et. al.}~\cite{PhysRevLett.129.187602}, originally derived purely from symmetry considerations. 
Here, motif connectivity can be mediated by inversion $\mathcal{P}$, a twofold rotation $R_{2}\parallel\boldsymbol{c}$, a mirror plane $\sigma\perp\boldsymbol{c}$, or an improper rotation $S_{n}\parallel\boldsymbol{c}$. 
For instance, when motifs are related by $\mathcal{P}$, the magnetoelectric is
\begin{equation}
\mathcal{H}_{\text{ME}}(\boldsymbol{k}) = -2\mu_{\text{B}} \eta_0\lambda_{1} (\boldsymbol{\sigma}\cdot\boldsymbol{m})\, (\boldsymbol{d}\cdot\boldsymbol{k}),
\end{equation}
while quadrupolar contributions vanish, placing such systems outside the altermagnetism category.
Notably, as discussed by Duan et. al.~\cite{PhysRevLett.134.106801}, achieving controllable SS requires motif connectivity mediated by $R_2$ rotations perpendicular to the local polar axis, defining a distinct subclass within altermagnetic materials. Although the dipolar term dominates in this regime, symmetry may still permit subleading high-order electric multipoles corrections.

In reorienting symmetry classes (see Table~II), electric dipoles at different sites are canted relative to each other, generating a net electric polarization (third line in Figure~\ref{Fig2}). 
This mechanism is only symmetry-allowed in NCS polar crystals. 
In the magnetoelectric coupling, antiparallel (parallel) dipole components across magnetic motifs add constructively (destructively). 
For instance, for dipoles $\boldsymbol{d}_{\pm} = (\pm d_x, \pm d_y, d_z)$, the resulting polarization points along $z$, while the induced SS remains oriented along $k_x$ and $k_y$. 
Thus, a net polarization, e.g., generated by an external electric field, can induce SS, enabling SP reversal as observed in a few reported cases~\cite{PhysRevLett.134.106802}.

\section{Monopolar Spin Splitting and $k$-Independent Effects at $\Gamma$}
Having addressed quadrupolar and dipolar contributions, we now examine the SS arising from monopolar magnetoelectric term (fourth line in Figure~\ref{Fig2}). 
Unlike dipole- or quadrupole-induced SS, which inherently depend on $\boldsymbol{k}$, the monopolar contribution yields a $k$-independent SS, allowing SS even at the high-symmetry $\Gamma$ point. 
In an electrically neutral unit cell, monopolar contributions from distinct motifs must globally cancel, i.e., $Q_n = (-1)^{n-1}Q_{0}$ (with $n=1,2$), leading to
\begin{equation}
    \mathcal{H}_{\text{ME}} = -2\mu_{\text{B}}\eta_0\lambda_0 Q_{0}\, \boldsymbol{\sigma}\cdot\boldsymbol{m},
\end{equation}
a condition achievable by doping within motifs without altering the magnetic moments, e.g., MnSiSnN$_{4}$~\cite{PhysRevLett.133.216701}. 
Although the monopolar term dominates the $k$-independent response, symmetry generally permits higher-order multipoles, which may introduce additional $k$-dependent features superimposed on the constant splitting, as observed in MnSiSnN$_{4}$~\cite{PhysRevLett.133.216701}.

Remarkably, even in systems with zero net magnetization, $\mathcal{H}_{\text{ME}}$ mirrors the functional form of the Zeeman SS, however, similar to the Zeeman effect, it fails to determine the energetic scale of the SS. 
The magnetization generates a Zeeman splitting of only few meV, two orders of magnitude smaller than the observed exchange splitting in ferromagnets\cite{sciadv.abn1401,PhysRevLett.88.047201,Norden2019}. 
Thus, while both conventional Zeeman effect and magnetoelectric terms correctly describe the SP and its symmetry properties, they substantially underestimate the splitting magnitude, requiring an effective exchange field far exceeding the magnetization itself. This holds not only for monopolar contributions, but also for higher-order multipoles in $\mathcal{V}_n$, which may coexist with charge disproportionation. As a result, spin splitting at $\Gamma$ is typically accompanied by a $k$-dependent component~\cite{PhysRevLett.133.216701}. These conclusions naturally extend to ferromagnets, where all magnetic motifs align and multipolar contributions add constructively, i.e., the relevant term in ferromagnets scales with their \textit{sum}. 

For motifs with identical electric potentials $\mathcal{V}_n$ and magnetic moments $\boldsymbol{m}_n = \boldsymbol{m}$, the magnetoelectric correction reads $\mathcal{H}_{\text{ME}} = -\mu_{\text{B}}\eta_{0}(\mathcal{V}_1 + \mathcal{V}_2)\boldsymbol{\sigma}\cdot\boldsymbol{m}$. 
The physical implications differ strikingly from the compensated case:  
\textit{i}) A nonzero monopolar term ($Q_n\neq0$) enhances the $k$-independent SS at $\Gamma$, augmenting the conventional Zeeman term, \textit{ii}) dipolar terms, aligned across motifs, induce a linear-in-$\boldsymbol{k}$ SS even in the absence of $\mathcal{P}$-breaking, which add up constructively (destructively) for parallel (antiparallel) dipole components, and \textit{iii}) quadrupolar terms, now additive, generate $k$-quadratic SS with enhanced anisotropy compared to the compensated case, particularly when motifs share lower symmetries with a single mirror, the SS is proportional to $(\boldsymbol{\sigma}\cdot\boldsymbol{m})\left[\mathcal{Q}_{x}(k^{2}_{x}+k^{2}_{y})-2\mathcal{Q}_{x}k_{z}^{2}\right]$.
The manipulation of the magnetic or dipole moments through external fields can thus potentially provide a SS control mechanism.

Fully compensated ferrimagnetic materials provide another prototypical setting in which $\mathcal{H}_{\text{ME}}$ plays a central role. In these systems, local magnetic moments at symmetry-inequivalent sites—often associated with chemically distinct elements such as 3$d$ and 4$d$ transition metals—can be antiparallel and cancel exactly, yielding zero net magnetization. However, the inequivalence of the underlying atomic environments ensures distinct electric potentials $\mathcal{V}_n$ at each site, resulting in a nonvanishing magnetoelectric correction. Consequently, all electric multipoles are generally symmetry-allowed, and the resulting SS includes not only a $k$-independent contribution at $\Gamma$, but also multipolar $k$-dependent features such as linear-in-$k$ and quadratic-in-$k$ terms as have been observed in ferrimagnets prototypes~\cite{PhysRevLett.134.116703,Guo2025,arxiv.2412.17232}, in complete analogy with ferromagnets. 
While in these systems, SS is similar to ferromagnetic SS, their compensated magnetic nature offers a unique testbed for our approach, which naturally captures their behavior through local motif asymmetries in the electric potential.
The presence of multiple active multipoles makes compensated ferrimagnets a fertile platform for probing the full hierarchy of symmetry-allowed SS effects within the framework proposed here.

\section{Conclusion}
This work uncovers a fundamental magnetoelectric mechanism governing SS across magnetic materials. We show that local electric multipoles coupled to magnetic moments drive distinct SS behaviors even without SOC. Our findings not only reconcile previous symmetry-based approaches but also predict characteristic $k$-dependence, opening new directions for engineering spin phenomena via local symmetry and electric ordering. This unified framework establishes magnetoelectric coupling as a key principle in spintronics beyond the traditional Zeeman and Rashba paradigms. 

The illustrative examples of $\Gamma$-splitting, spin Zeeman splitting, and altermagnetism correspond to dominant monopolar, dipolar, and quadrupolar contributions in the local electric potential, yet the formalism is general and extends to arbitrary multipolar orders. In particular, the explicit expression 
\[
\mathcal{H}_{\text{ME}} = -\mu_{\text{B}}\eta_0(\mathcal{V}_1 - \mathcal{V}_2)\boldsymbol{\sigma} \cdot \boldsymbol{m}
\]
clarifies how SS arises from symmetry-allowed differences in the scalar potential across magnetic motifs connected by crystallographic operations. 
The framework also naturally accounts for fully compensated ferrimagnets, where antiparallel moments residing on chemically distinct sites give rise to SS at $\Gamma$ through local electric potential differences. In such systems, all electric multipoles are symmetry-allowed, leading to coexisting $k$-independent and $k$-dependent spin splitting. 
Our analysis shows that the absence of SOC does not preclude the existence of relativistic corrections, as captured by DFT calculations. The magnetoelectric coupling derived here emerges from the Dirac formalism independently of SOC, highlighting that the SS in compensated magnets can be interpreted as a relativistic correction that persists even in the absence of conventional SOC. 
The magnetoelectric term directly encodes the role of motif connectivity and site asymmetry, and enables the classification of all SS-permitting crystal symmetries in terms of their associated $k$-dependencies. The proposed relativistic correction establishes a missing link in the theoretical foundation of spintronic effects and provides a predictive framework to design SS in magnets.

\begin{acknowledgments}
The author acknowledges insightful discussions with Alex Zunger and Linding Yuan on spin splitting in antiferromagnetic materials, and with José-Antonio Sanchez and Luis Henrique Mera on the magnetoelectric origin of spin splitting in magnets. 
Computational support was provided by the \textit{Central Computacional Multiusuário} of UFABC. 
This work was supported by the São Paulo Research Foundation (FAPESP) under Grant No. 2023/09820-2.

\end{acknowledgments}


%

\end{document}